\documentclass[journal=jpclcd,manuscript=letter]{achemso}
\usepackage{graphicx}
\usepackage{amsfonts}
\usepackage{amsmath}
\usepackage{amssymb}
\usepackage{float}

\title{Isotope effects in the electronic spectra of ammonia from ab initio
semiclassical dynamics}
\author{\={E}riks Kl\={e}tnieks}
\affiliation[EPFL]{Laboratory of Theoretical Physical Chemistry, Institut des Sciences et Ing\'enierie Chimiques, Ecole Polytechnique F\'ed\'erale de Lausanne (EPFL), CH-1015, Lausanne, Switzerland}
\author{Yannick Calvino Alonso}
\author{Ji\v{r}\'i J.L. Van\'i\v{c}ek}
\affiliation[EPFL]{Laboratory of Theoretical Physical Chemistry, Institut des Sciences et Ing\'enierie Chimiques, Ecole Polytechnique F\'ed\'erale de Lausanne (EPFL), CH-1015, Lausanne, Switzerland}
\email{jiri.vanicek@epfl.ch}
\affiliation{Laboratory of Theoretical Physical Chemistry, Institut des Sciences et
Ing\'enierie Chimiques, Ecole Polytechnique F\'ed\'erale de Lausanne (EPFL),
CH-1015, Lausanne, Switzerland}

\graphicspath{{./Figures/}}

\begin{document}

\begin{tocentry}
\includegraphics{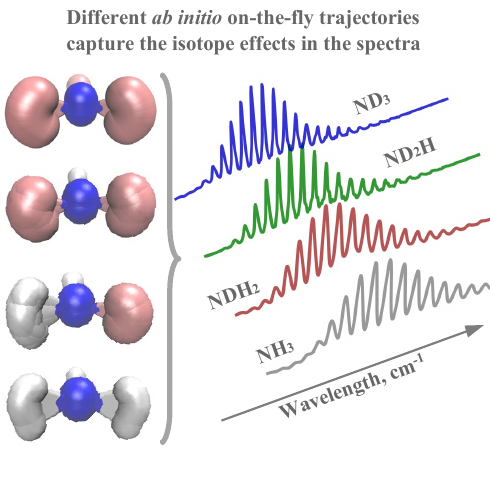}
\end{tocentry}

\begin{abstract}
Despite its simplicity, the single-trajectory thawed Gaussian approximation
has proven useful for calculating vibrationally resolved electronic spectra
of molecules with weakly anharmonic potential energy surfaces. Here, we show
that the thawed Gaussian approximation can capture surprisingly well even
more subtle observables, such as the isotope effects in the absorption
spectra, and we demonstrate it on the four isotopologues of ammonia (NH$_{3}$, NDH$_{2}$, ND$_{2}$H, ND$_{3}$). The differences in their computed spectra are due to the differences in the semiclassical trajectories followed by the four isotopologues, and the isotope effects---narrowing of the transition
band and reduction of the peak spacing---are accurately described by this
semiclassical method. In contrast, the adiabatic harmonic model shows a
double progression instead of the single progression seen in the
experimental spectra. The vertical harmonic model correctly shows only a single progression but fails to describe the anharmonic peak spacing. Analysis of the normal-mode activation upon excitation provides insight into the elusiveness of the
symmetric stretching progression in the spectra.
\end{abstract}

Vibrationally resolved electronic spectroscopy has made significant
contributions to the understanding of the structure and dynamics of
molecules. However, extracting information about the potential energy
surface (PES) on which the dynamics occur remains challenging due to several
factors influencing the experimentally observed spectra. Since the electronic structure is invariant to the isotope substitution, measuring the isotope effects in the spectra not only provides additional information about the shape of the surface and dynamics, but also aids in the assignment of transitions. Computational methods help to interpret experimental results and
provide a clearer understanding of the dynamics responsible for the
observed spectra. \par

Within the Born-Oppenheimer approximation,\cite{Born_Oppenheimer:1927,
book_Heller:2018} the isotope effects in the spectra are attributed to the
changes in the characteristic frequencies of isotopologues, which depend
strongly on the mass of the substituted atom. Depending on the system of
interest, the isotope exchange can shift the peak positions to higher or
lower energies. In general, the isotope substitution changes the energy of
the 0-0 transition, the width of the transition band, and also the spacing, widths, and intensities of peaks. The magnitude of these changes depends on the
mass ratio of the substituted species and on the displacement in the
vibrational normal modes responsible for the spectra.\cite%
{book_Wilson_Cross:1980, book_Harris_Bertolucci:1989}

Among the various methods for calculating the isotope effects in the
spectra, the simplest approach is based on the global harmonic approximation
to the potential.\cite{Hazra_Nooijen:2005, AvilaFerrer_Santoro:2012,
Biczysko_Barone:2009, Baiardi_Barone:2013, Cerezo_Santoro:2013} Harmonic
models are computationally cheap and can capture some isotope effects
correctly but fail in systems with a large-amplitude nuclear motion.\cite%
{Begusic_Vanicek:2022} In contrast, the exact quantum dynamics on a grid\cite%
{Kosloff:1988} or multiconfigurational time-dependent Hartree (MCTDH) method%
\cite{Beck_Jackle:2000} reproduce all of the isotope effects seen in the
experimental spectra, but at the high cost of constructing a global PES.\cite%
{Marquardt_Quack:2011} A compromise is offered by the semiclassical
trajectory-based methods, which, in terms of accuracy, lie between the
harmonic and exact results.\cite{Tatchen_Pollak:2009, Wehrle_Vanicek:2014,
Wehrle_Vanicek:2015, Gabas_Ceotto:2017, Gabas_Ceotto:2018,
Gabas_Ceotto:2019, Bonfanti_Pollak:2018, Patoz_Vanicek:2018,
Begusic_Vanicek:2018, Begusic_Vanicek:2020, Golubev_Vanicek:2020,
Prlj_Vanicek:2020, Botti_Conte:2021}

The single-trajectory thawed Gaussian approximation (TGA), introduced by
Heller,\cite{Heller:1975} is a semiclassical method, which is accurate for
short propagation times and which has been used to calculate both absorption
and emission spectra of weakly anharmonic molecules with a large number of
degrees of freedom.\cite{Grossmann:2006, Wehrle_Vanicek:2014,
Wehrle_Vanicek:2015} In weakly anharmonic systems and for short times
relevant in electronic spectroscopy, the single Gaussian wavepacket used in
the TGA is, sometimes surprisingly, sufficient to sample the dynamically
important region of the phase space. Moreover, the classical trajectory
associated with the wavefunction provides a simplified, intuitive picture of the
dynamics, while the evolving width of the Gaussian wavepacket partially
captures the quantum effects. Successful past applications of the TGA
prompted us to apply it to the isotope effects. Although the PES used for
the propagation is invariant under the isotope substitution, for each
isotopologue, the guiding trajectory explores a different region of this
common PES. This requires propagating a new trajectory for each
isotopologue, but the total computational cost is still considerably lower
than the cost of constructing the full surface.

To investigate the isotope effect on the spectrum, a comprehensive
understanding of the PES associated with the transition is essential. In the
case of ammonia, the first excited-state $\tilde{A}$ is quasi-bound in the
Franck-Condon region.\cite{Li_Vidal:1994, Chung_Ziegler:1988} A finite
barrier separates the bound region from the conical intersection of the $%
\tilde{X}$ and $\tilde{A}$ states. This conical intersection couples the two
surfaces nonadiabatically and is responsible for an internal conversion,
which leads to the broadening of the absorption spectra.\cite{Seideman:1995,
Lai_Guo:2008, Ma_Guo:2014} The lifetime in the quasi-bound region depends on
the isotopologue and ranges from a few hundred femtoseconds to a few
picoseconds, allowing for more than several oscillations to occur before the
escape. The ammonia absorption spectra $\tilde{A}^{1}A_{2}^{\prime\prime}
\longleftarrow \tilde{X}^{1}A_{1}^{\prime}$ ($S_{1} \longleftarrow S_{0}$)
has a long progression that is induced by the activation of the symmetric
bending (umbrella motion) and symmetric stretching modes.\cite%
{Chen_Caldwell:1999, Burton_Meath:1993,Tang_Tannor:1990} Although the
experimental spectra of isotopologues (NDH$_{2}$, ND$_{2}$H, ND$_{3}$) have
similar patterns as the spectrum of NH$_{3}$, the isotope effects are
clearly visible.\cite{Cheng_Yung:2006} As the number of hydrogen atoms
substituted by deuterium increases, the energy of the 0-0 transition
increases, while the peak spacing and width decreases. In addition, the
transition band becomes narrower.

The isotope substitution can also affect the molecular symmetry. Although
the ground-state geometry of all four ammonia isotopologues has a pyramidal
shape, the NH$_{3}$ and ND$_{3}$ belong to the C$_{3v}$ point group, whereas
NDH$_{2}$ and ND$_{2}$H belong to the C$_{s}$ point group. As the symmetry
changes from one isotopologue to another, the dynamics on the excited-state
surface changes not only due to the change of reduced masses, but also due
to the activation of other normal modes.\cite{book_Harris_Bertolucci:1989}

\begin{figure}
\centering
\includegraphics{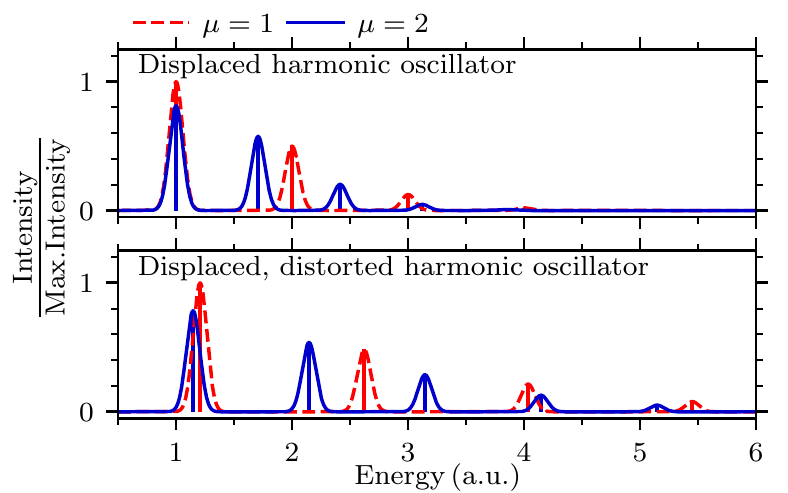} 
\caption{Isotope effects in the
electronic spectra of one-dimensional harmonic oscillators. The finite-width
peaks were obtained with the TGA, whereas the vertical lines indicate the
spectra obtained with the time-independent approach,\cite{book_Tannor:2007,
Heller:1981a} in which the Franck-Condon factors were calculated with
Eq.~(\ref{eq:FC_factors_DHO}) (in the top panel) or Eq.~(3.26) from
Ref.~\citenum{Iachello_Ibrahim:1998} (in the bottom panel). The spectra
were scaled according to the maximum intensity of the system with $\mu=1$.
The details of this calculation can be found in the Supporting Information.}
\label{fig:isotope_effects_example}
\end{figure}

To demonstrate various isotope effects on electronic spectra, let us first
consider an analytically solvable one-dimensional displaced harmonic
oscillator model. In this model, the two PESs involved in the transition are
described by harmonic oscillators with the same force constant ($%
k_{g}=k_{e}=k$), but whose minima are displaced horizontally by $\Delta q$
and vertically by an energy gap $\Delta E$. Assuming that the initial state
is the vibrational ground state $|\Psi _{g,0}\rangle $ of the electronic
ground state, the Franck-Condon factors, which determine the transition
probability between two vibronic states and hence the intensities of the
spectral peaks, follow the Poisson distribution 
\begin{equation}
|\langle \Psi _{g,0}|\Psi _{e,n}\rangle |^{2}=e^{-S_{ge}}\frac{S_{ge}^{n}}{n!%
},  \label{eq:FC_factors_DHO}
\end{equation}%
where $S_{ge}=\Delta q^{2}\frac{\mu \omega _{g}}{2\hbar }$ is the Huang-Rhys
parameter for the ground and excited states, $\omega
_{g}=(k_{g}/\mu )^{1/2}$, and $\mu $ is the reduced mass of the oscillator.
Analytical expressions for Franck-Condon factors for squeezed and displaced
harmonic oscillators (in which $k_{e}\neq k_{g}$) can be found, e.g., in Ref.~%
\citenum{Iachello_Ibrahim:1998}. The spectra computed in these model systems with the time-independent
approach are compared with the results obtained with the TGA, which is exact
in quadratic potentials, in Fig.~\ref{fig:isotope_effects_example}. In spite
of the simplicity of the models, the spectra exhibit all of the
above-mentioned isotope effects---namely the changes in the 0-0 transition, peak
spacing, width of the transition band---except for the effect on the peak
widths, which is zero in the time-independent approach and arbitrary in the
TGA (due to the choice of the damping of the wavepacket time autocorrelation
function). Note that the 0-0 transition is only affected in the distorted
model, where the vibrational frequencies of the ground and excited states
differ.

Polyatomic molecules with at least two vibrational degrees of freedom pose
an additional challenge due to the multidimensional nature of their
potential energy surfaces. Within the harmonic approximation, the
excited-state potential energy surface $V(q)$ of a molecule is
approximated by a quadratic expansion about the reference geometry $q_{\text{%
ref}}$: 
\begin{equation}
V_{\text{HA}}(q)=V|_{q_{\text{ref}}}+V^{\prime }|_{q_{\text{ref}%
}}^{T}\cdot x_{r}+x_{r}^{T}\cdot V^{\prime \prime }|_{q_{\text{%
ref}}}\cdot x_{r}/2,  \label{eq:HA_general_equation}
\end{equation}%
where $x_{r}:=q-q_{\text{ref}}$, $V^{\prime }|_{q_{\text{ref}}}=\text{grad}%
_{q}V|_{q_{\text{ref}}}$ is the gradient vector, and $V^{\prime \prime }|_{q_{%
\text{ref}}}=\text{Hess}_{q}V|_{q_{\text{ref}}}$ is the Hessian matrix at
the reference geometry. The PES of the excited state is generally not only
displaced and distorted but also rotated with respect to the surface of the
initial state. This rotation is called the Duschinsky effect and is
characterized by the Duschinsky matrix $J$ that relates the normal-mode
coordinates of the two states.\cite{Duschinsky:1937} The general form of the
potential in Eq.~(\ref{eq:HA_general_equation}) allows for a free choice of $%
q_{\text{ref}}$, but two choices are the most natural. The adiabatic
harmonic model is constructed about the equilibrium geometry of the
excited-state surface, whereas the vertical harmonic model expands the
excited-state surface about the Franck-Condon geometry, i.e., the
equilibrium geometry of the ground-state surface.\cite%
{AvilaFerrer_Santoro:2012,Sattasathuchana_Baldridge:2020} Various exact and
efficient algorithms are able to treat global harmonic models of large
systems, and the thawed Gaussian approximation is one of them. The harmonic
models are easy to construct, as they only require a single Hessian
calculation in the ground and excited states once the geometries have been
optimized. The subsequent evaluation of the autocorrelation functions and
spectra requires much less computational effort. The drawback of the
harmonic models is the complete neglect of anharmonicity, which may lead to
erroneous results for systems with a large amplitude nuclear motion.

Going beyond the harmonic approximation requires the inclusion of
anharmonicity effects, and Heller's\cite{Heller:1975} thawed Gaussian
approximation does it at least partially by employing the local harmonic
approximation 
\begin{align}
V_{\text{LHA}}(q;q_{t}) = & V|_{q_{t}}+V^{\prime }|_{q_{t}}^{T}\cdot x_{t}\nonumber \\
& + x_{t}^{T}\cdot V^{\prime \prime }|_{q_{t}}\cdot x_{t}/2
\end{align}%
of the potential about the current center $q_{t}$ of the wavepacket ($%
x_{t}:=q-q_{t}$). The TGA allows the exploration of anharmonic parts of the
potential without the need to construct a full PES \textit{a priori}. Within
the time-dependent approach to spectroscopy,\cite{Vanicek_Begusic:2021,
book_Tannor:2007} the nuclear wavepacket is propagated by solving the
time-dependent Schr{\"{o}}dinger equation 
\begin{equation}
i\hbar |\Dot{\Psi}_{t}\rangle =\hat{H}|\Psi _{t}\rangle ,
\label{eq:Schrodinger_equation}
\end{equation}%
where $|\Psi _{t}\rangle $ is the nuclear wavepacket, $\hat{H}:=\Hat{p}%
^{T}\cdot m^{-1}\cdot \Hat{p}/2+V(\Hat{q})$ is the
vibrational Hamiltonian, and ~$m=\text{diag}(m_{1},...,m_{D})$ is the
diagonal mass-matrix. The thawed Gaussian approximation assumes that the
nuclear wavepacket is a $D$-dimensional Gaussian,\cite%
{book_Lubich:2008,Lasser_Lubich:2020} which, in Hagedorn's parametrization,%
\cite{Hagedorn:1980,Hagedorn:1998} is written as 
\begin{align}
\psi (q,t)= N_{t}\exp \Big [\frac{i}{\hbar }\Big ( & \frac{1}{2} x_{t}^{T}\cdot P_{t}\cdot Q_{t}^{-1}\cdot x_{t}\nonumber \\
& + p_{t}^{T}\cdot x_{t}+S_{t} \Big )\Big ],\label{eq:Gaussian_WP}
\end{align}%
where $N_{t}=(\pi \hbar )^{-D/4}(\text{det}Q_{t})^{-1/2}$ is the
normalization constant, $x_{t}:=q-q_{t}$ is the shifted position,
$q_{t}$ and $p_{t}$ are the position and momentum of
the wavepacket's center, $S_{t}$ is the classical action, and $P_{t}$, $%
Q_{t} $ are complex $D\times D$ matrices, which satisfy the relations 
\begin{align}
Q_{t}^{T}\cdot P_{t}-P_{t}\cdot Q_{t}& =0, \\
Q_{t}^{\dagger }\cdot P_{t}-P_{t}^{\dagger }\cdot Q_{t}& =2iI_{D},
\end{align}%
in which $I_{D}$ is the $D\times D$ identity matrix. Approximating $V$ with $%
V_{\text{LHA}}$ in the Schr{\"{o}}dinger Eq.~(\ref{eq:Schrodinger_equation})
yields the nonlinear Schr{\"{o}}dinger equation%
\begin{equation*}
i\hbar |\Dot{\Psi}_{t}\rangle = [\Hat{p}^{T}\cdot m^{-1}\cdot 
\Hat{p}/2+V(\Hat{q};q_{t}) ]|\Psi _{t}\rangle .
\end{equation*}%
This equation is solved exactly by the Gaussian ansatz if the Gaussian's
parameters satisfy the first-order differential equations 
\begin{align}
\Dot{q}_{t}& =m^{-1}\cdot p_{t},  \label{eq:q_prop} \\
\Dot{p}_{t}& =-\text{grad}_{q}V|_{q^{t}},  \label{eq:p_prop} \\
\Dot{Q}_{t}& =m^{-1}\cdot P_{t},  \label{eq:Q_prop} \\
\Dot{P}_{t}& =-\text{Hess}_{q}V|_{q^{t}}\cdot Q_{t},  \label{eq:P_prop} \\
\Dot{S}_{t}& =L_{t},  \label{eq:S_prop}
\end{align}%
where $L_{t}$ denotes the Lagrangian 
\begin{align}
L_{t}& =\Dot{q}_{t}^{T}\cdot m\cdot \Dot{q}_{t}/2-V(q_{t})  \notag \\
& =p_{t}^{T}\cdot m^{-1}\cdot p_{t}/2-V(q_{t}).
\end{align}%
Equations~(\ref{eq:q_prop}) and (\ref{eq:p_prop}) imply that the center of
the wavepacket follows exactly the classical trajectory of the original
potential, while the width of the Gaussian is propagated with the effective
potential $V_{\text{LHA}}(q_{t})$ [see Eqs.~(\ref{eq:Q_prop}) and (\ref%
{eq:P_prop})].

Being a single-trajectory method, the TGA can be easily combined with an
on-the-fly evaluation of the electronic structure. Most often, electronic
structure calculations for a molecule with $N$ atoms are performed in $3N$
Cartesian coordinates. However, natural coordinates for the propagation of
the thawed Gaussian wavepacket and for the construction of global harmonic
models are the $3N-6$ mass-scaled vibrational normal-mode coordinates $q_{j}$%
. We perform the propagation in the excited-state vibrational normal-mode
coordinates, since these coordinates provide the most natural description of
the dynamics following the electronic excitation. The transformation from
the Cartesian to normal-mode coordinates requires the removal of
translational and rotational degrees of freedom. Although the vibrations and
rotations are not fully separable, we reduce the coupling by translating and
rotating the nuclei of the molecule into the Eckart frame. Here, we closely
follow the procedure described in more detail in Ref.~%
\citenum{Vanicek_Begusic:2021}.

Let us denote the full molecular configuration as $\xi:=(\mathbf{r}_{1},
\dots, \mathbf{r}_{N})$, where $\mathbf{r}_{a}$ are the Cartesian
coordinates of atom $a$. We introduce a reference molecular configuration $%
\xi_{\text{ref}}:=(\mathbf{r}_{\text{ref},1}, \dots, \mathbf{r}_{\text{ref}%
,N})$. In all our calculations $\xi_{\text{ref}}$ is the equilibrium
geometry on the PES of the excited state $\tilde{A}$ of ammonia. First, the
translational degrees of freedom are removed by shifting coordinates of each
atom to the center-of-mass frame: 
\begin{equation}
\mathbf{r} _{a}^{\prime}:=\mathbf{r} _{a} - \mathbf{r}_{\text{com}},
\label{eq:center_of_mass}
\end{equation}
where $\mathbf{r}_{\text{com}}=(\sum_{a=1}^{N}m_{a} \mathbf{r}%
_{a})/(\sum_{a=1}^{N}m_{a})$ and $m_{a}$ is the mass of the nucleus of atom $%
a$. The translated molecular configuration is $\xi^{\prime}:=(\mathbf{r}%
^{\prime}_{1}, \dots, \mathbf{r}^{\prime}_{N})$. In the following, we assume
that the center of mass of the reference configuration $\xi_{\text{ref}}$ is
at the origin, i.e., $\mathbf{r}_{\text{ref,com}}=0$.

In the second step, we minimize the rovibrational coupling by rotating the
configuration $\xi^{\prime}$ into the Eckart frame. This is equivalent to
minimizing the squared mass-scaled distance\cite{Kudin_Dymarsky:2005} 
\begin{equation}
||\xi_{\text{ref}}-\xi_{\text{rot}}||^{2}:=\sum_{a=1}^{N}m_{a}|\mathbf{r}_{%
\text{ref,a}}-\mathbf{r}_{\text{rot,a}}|^{2},  \label{eq:xyz_rotation}
\end{equation}%
between the rotated configuration $\xi_{\text{rot}}$ and the reference
configuration $\xi_{\text{ref}}$. Here $\mathbf{r}_{\text{rot,a}}:=R \cdot 
\mathbf{r}^{\prime}_{a}$. The required $3 \times 3$ rotation matrix $R$ can
be found, e.g., by the Kabsch algorithm.\cite{Kabsch:1978} The
transformation to and from normal-mode coordinates is based on the
orthogonal matrix $O_{\text{ref}}$ that diagonalizes the mass-scaled
Cartesian Hessian matrix at $\xi_{\text{ref}}$ on the excited-state surface $%
V$:
\begin{equation}
O_{\text{ref}}^{T}\cdot m^{-1/2}\cdot \text{Hess}_{\xi}V|_{\xi _{\text{ref}%
}}\cdot m^{-1/2}\cdot O_{\text{ref}}=\Omega^{2},
\label{eq:hessian_diagonalization}
\end{equation}%
where $m$ and $\Omega$ are $3N \times 3N$ diagonal matrices, containing,
respectively the values of the atomic masses and normal-mode frequencies
(still including the zero frequencies). After projecting out the
zero-frequency modes associated with rotations and translations, the overall
transformation from the Cartesian coordinates $\xi$ to the mass-scaled
vibrational normal-mode coordinates $q$ is 
\begin{equation}
q =L_{\text{ref}}^{T}\cdot m^{\frac{1}{2}}\cdot (R_{\xi } \cdot \xi^{\prime}
- \xi_{\text{ref}}),  \label{eq:q_transform}
\end{equation}
where $L_{\text{ref}}$ is the leading $3N\times(3N-6)$ submatrix of $O_{%
\text{ref}}$, and $R_{\xi }=I_{N}\otimes R$ is a $3N\times 3N$
block-diagonal matrix, whose $N$ $3\times3$ blocks are identical and equal
to the rotation matrix $R$. Similarly, we can obtain the potential gradient
and Hessian in vibrational normal-mode coordinates: 
\begin{align}
\text{grad}_{q}V& =L_{\text{ref}}^{T}\cdot m^{-\frac{1}{2}}\cdot R_{\xi
}\cdot \text{grad}_{\xi},  \label{eq:gradq_transformation} \\
\text{Hess}_{q}V& =L_{\text{ref}}^{T}\cdot m^{-\frac{1}{2}}\cdot R_{\xi
}\cdot \text{Hess}_{\xi}\cdot R_{\xi }^{T}\cdot m^{-\frac{1}{2}}\cdot L_{%
\text{ref}}.  \label{eq:Hessian_transformation}
\end{align}%
Equation~(\ref{eq:q_transform}) can be rearranged to transform the
normal-mode coordinates back to the Cartesian coordinates as 
\begin{equation}
\xi_{\text{rot}} =m^{\text{-}\frac{1}{2}}\cdot L_{\text{ref}}\cdot q + \xi _{%
\text{ref}}.  \label{eq:S_q2xyz}
\end{equation}
The above framework allows combining \textit{ab initio} calculations in
Cartesian coordinates with the propagation of the thawed Gaussian wavepacket
in normal-mode coordinates. Equations~(\ref{eq:q_transform})--(\ref%
{eq:Hessian_transformation}) are also used for the construction of global
harmonic models.

In what follows, all \textit{ab initio} calculations of ammonia were
performed using the complete active-space second-order perturbation theory
CASPT2(8/8) method with Dunnings correlation-consistent basis set \textit{%
aug-cc-pVTZ}; a level shift of 0.2 a.u. was applied to avoid the
intruder-state problem. \cite{Celani_Werner:2003, Ross_Andersson:1995} All
of the calculations were performed with Molpro2019 package.\cite{MOLPRO:2019}
The Gaussian wavepacket was propagated with a time step of 8 a.u. for 1000
steps (1 a.u. $\approx $ 0.024 fs resulting in a total time of 8000 a.u. $%
\approx 193.5$ fs) using the second-order symplectic integrator.\cite%
{Vanicek:2023} To remove the systematic errors of the \textit{ab initio}
vertical excitation energies, all computed spectra in Fig.~\ref%
{fig:all_spectra} were shifted independently to obtain the best fit to the
experiment. In addition, the spectra were broadened by a Lorentzian with
half-width-at-half-maximum of 170, 137, 95, and 110 cm$^{-1}$
for NH$_{3}$, NDH$_{2}$, ND$_{2}$H, and ND$_{3}$, respectively. Neither the
overall shift nor the broadening affect the subsequent analysis of peak
spacing, peak intensities, and width of the spectral envelope, which are
independent of the shift and broadening. Additional details can be found in
the Supporting Information.

The initial Gaussian wavepacket was the ground vibrational eigenstate of the
harmonic fit to the PES of the ground electronic state ($\tilde{X}%
^{1}A_{1}^{\prime}$) at one of the two degenerate minima. The initial
position $q_{0}$ in normal-mode coordinates was obtained from Eq.~(\ref%
{eq:q_transform}), where $\xi^{\prime}$ is replaced with the Cartesian
coordinates $\xi_{\text{init}}$ corresponding to the equilibrium geometry of
the ground state. The initial momentum $p_{0}$ of the wavepacket was zero.
The initial $Q_{0}$ and $P_{0}$ matrices, which control the width of the
Gaussian wavepacket, were 
\begin{equation}
P_{0}  = i \hbar \Gamma^{\frac{1}{2}} \quad \text{and} \quad Q_{0}  = \hbar \Gamma^{-\frac{1}{2}},  \label{eq:Q_0_P_0}
\end{equation}
where $\Gamma:=\text{Hess}_{q}V^{(g)}|_{q_{0}}$ is the Hessian in
normal-mode coordinates calculated at the equilibrium geometry of the ground
state.

\begin{figure*}
\centering
\includegraphics{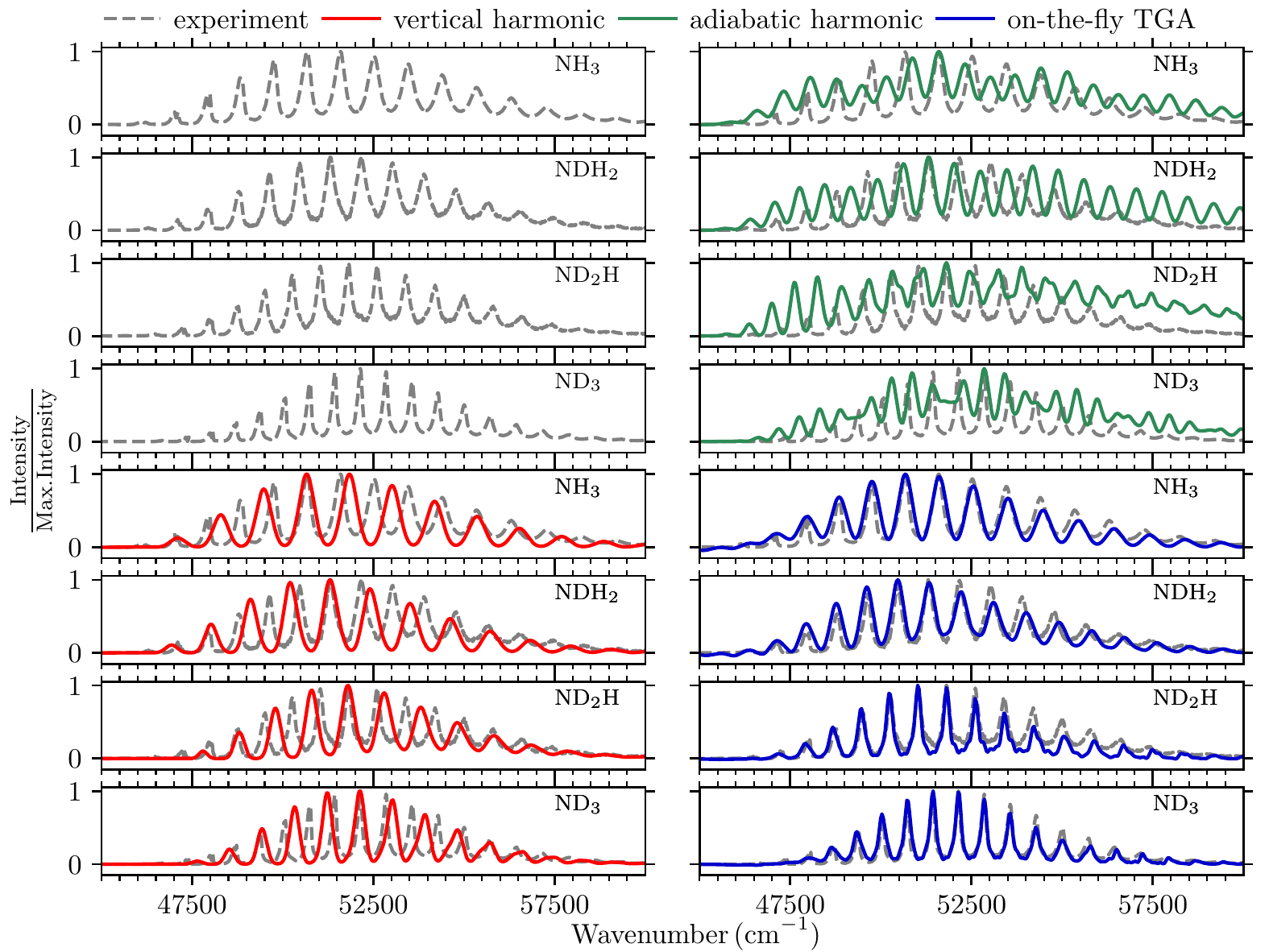} 
\caption{Simulated electronic
absorption spectra of ammonia isotopologues (NH$_{3}$, NDH$_{2}$, ND$_{2}$H,
and ND$_{3}$) are compared to the experimental spectra\cite{Cheng_Yung:2006}
recorded in the gas phase at 298 K.}\label{fig:all_spectra}
\end{figure*}

The experimental spectra of ammonia isotopologues, shown in the top left-hand
panel of Fig.~\ref{fig:all_spectra}, clearly exhibit the isotope effects on
the 0-0 transition energy, on the width of the transition band, and on the peak spacing, width, and intensity. First, the shift in the 0-0 transition
towards higher energies can be directly compared with the adiabatic
excitation energy from \textit{ab initio} calculations in Table~\ref%
{tab:00_transition_isotope_effects}, indicating that quantum chemical
calculations correctly reproduce the trend. Second, the bands associated to the $\tilde{A}%
^{1}A_{2}^{\prime\prime} \longleftarrow \tilde{X}^{1}A_{1}^{\prime}$
transition become narrower. Third, the decrease in the peak
spacing for highly substituted species is in line with the fact that the
spacing between the vibrational energy levels decreases with increasing
reduced mass. Lastly, the nonradiative processes, such as
tunnelling and internal conversion,\cite{Ma_Guo:2014} may cause the increase in the peak widths. Indeed, highly deuterated isotopologues have spectra with narrower peaks because they have a longer
lifetime in the quasi-bound region, which is due to the weaker tunneling for
heavier isotopes.

\begin{table}
\centering
\begin{tabular}{rrrr}
& NDH$_{2}$ & ND$_{2}$H & ND$_{3}$ \\ \hline\hline
Experiment & 171 & 347 & 532 \\ 
Adiabatic excitation energy & 128 & 260 & 399%
\end{tabular}
\caption{Isotope effect on the 0-0 transition of ammonia evaluated as the
difference $\Delta E_{00}(\text{ND}_{j}\text{H}_{i})-\Delta
E_{00}(\text{NH}_{3})$. The adiabatic excitation energy includes the
zero-point vibrational energy. All values shown are in cm$^{-1}$.} \label%
{tab:00_transition_isotope_effects}
\end{table}

The adiabatic harmonic model is constructed around the excited-state
equilibrium geometry, and the same Cartesian Hessian is used to construct
the PES for all isotopologues. The calculated spectra in the top right-hand panel
of Fig.~\ref{fig:all_spectra} show a double progression. In addition, for
all isotopologues the spectral envelope extends to higher energies,
reflecting the poor description of the short-time dynamics of the system.
Overall, this confirms that the adiabatic harmonic model is a bad
approximation of the strongly anharmonic PES of the ammonia molecule, where
the differences between ground- and excited-state equilibrium geometries are
significant. In the following, we do not discuss the isotope effect on the
0-0 transition, since the shift applied to compensate for the overall errors
of the electronic structure calculations (see Supporting Information) is of
the same order as the expected isotope effect. Instead, we focus on the
isotope effects that are independent of this shift.

As observed in Ref.~\citenum{Wehrle_Vanicek:2015}, the vertical harmonic
model in the case of NH$_{3}$ yields a single progression and recovers the
overall shape of the experimental spectrum. The results for other
isotopologues further confirm this observation. Moreover, the computed
envelopes of the spectra of all isotopologues agree rather well with the
envelopes of the experimental spectra. This is because the vertical harmonic
model approximates the PES well in the Franck-Condon region,
which determines the spectral envelope. The widths of spectral
envelopes are predicted rather accurately (see Table~\ref{tab:size_of_envelope}). The
model also describes qualitatively the trend of decreasing peak spacing.
However, as the model fails to capture anharmonicity, the peak spacing is
not reproduced quantitatively.

\begin{table}
\centering
\begin{tabular}{rllll}
& NH$_{3}$ & NDH$_{2}$ & ND$_{2}$H & ND$_{3}$ \\ \hline\hline
Exp. & 5554 & 5030 & 4700 & 4638 \\ 
TGA & 5426 & 5342 & 4792 & 4558 \\ 
vertical harmonic & 5466 & 5202 & 4808 & 4490 \\ 
adiabatic harmonic & 7044 & 7224 & 5540 & 5748%
\end{tabular}
\caption{Width of the spectral envelope in cm$^{-1}$. The width of the
spectral envelope was estimated as twice the square root of the weighted
variance of the peak positions, where the weight of a peak position is given
by its intensity (see the Supporting Information for details).} \label%
{tab:size_of_envelope}
\end{table}

Within the thawed Gaussian approximation, the classical trajectory guiding
the center of the wavepacket follows the exact and fully anharmonic \textit{%
ab initio} PES, whereas the width of the wavepacket feels anharmonicity only
approximately through the effective, locally harmonic potential. This
on-the-fly approach improves over both global harmonic models and recovers
well the isotope effects on the peak spacing and width of the spectral
envelope. The intensities of the peaks are reproduced very well near the 0-0
transition, whereas the differences are more pronounced at higher energies
for all isotopologues. The isotope effect on the peak spacing in Fig.~\ref%
{fig:peak_differences} shows a remarkable agreement with the experiment with
slight differences only near the 0-0 transition. In contrast, both harmonic
models deviate substantially from the linear dependency of the shift on the
peak frequency. The better performance of the on-the-fly TGA suggests that
the fully anharmonic classical trajectory employed in the TGA describes
better the true periodicity of the oscillations in the quasi-bound region of
the $\tilde{A}$ state. Although a new calculation must be performed for each
isotopologue, the results clearly indicate that this pays off.

\begin{figure}
\centering
\includegraphics[]{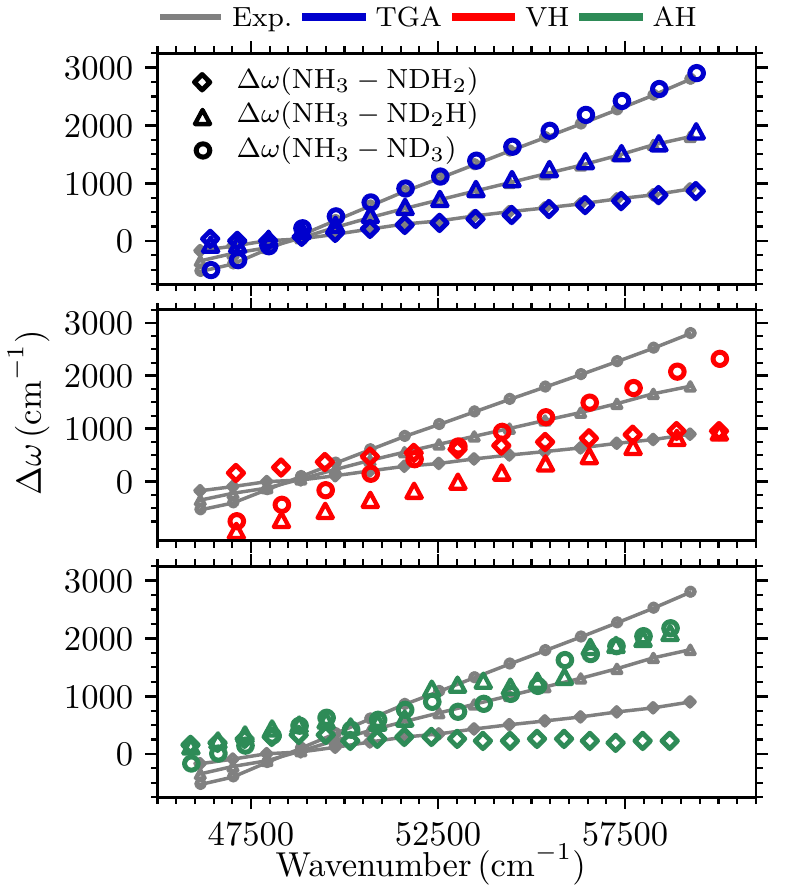} 
\caption{Isotope effects on the
peak spacing in the $\tilde{A}^{1}A_{2}^{\prime\prime} \longleftarrow
\tilde{X}^{1}A_{1}^{\prime}$ band computed with the on-the-fly thawed Gaussian
approximation (TGA), adiabatic harmonic (AH) model, and vertical harmonic
(VH) model are compared with the experimental values. $\Delta \omega$ is the
difference between the wavenumbers of corresponding peaks in the spectra of
two isotopologues.} \label{fig:peak_differences}
\end{figure}

One of the strengths of semiclassical methods is the ease with which they
reveal the molecular dynamics that generate these spectra. The thawed
Gaussian approximation makes this interpretation even simpler because it
relies on only one trajectory. Having performed all the calculations in the
excited-state normal-mode coordinates (depicted in Fig.~\ref%
{fig:nh3_vib_norm_modes}), we also know the time evolution of each of these
modes. In ammonia, the majority of the excited-state normal modes are
similar to the more commonly used ground-state normal modes. In the case of
NH$_{3}$ and ND$_{3}$, where the excited-state equilibrium geometry belongs
to the D$_{3h}$ point group, there are two pairs of degenerate normal
modes---a degenerate pair of asymmetric stretching modes and a degenerate
pair of scissoring modes. The other two modes---symmetric stretching and
symmetric bending (umbrella motion)--- are nondegenerate. In the case of
partially deuterated isotopologues (NDH$_{2}$ and ND$_{2}$H), where the
symmetry of the excited-state equilibrium geometry belongs to the C$_{2v}$
point group, all six normal modes are nondegenerate.

\begin{figure}
\centering
\includegraphics[]{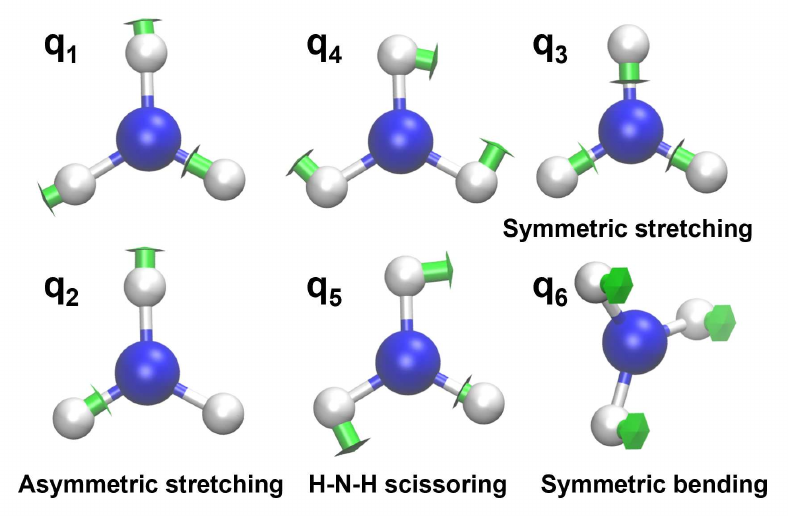} 
\caption{Vibrational normal modes of the
excited electronic state of ammonia. In NH$_{3}$ and ND$_{3}$ the two
asymmetric stretching modes ($q_{1}$ and $q_{2}$) are degenerate, and the
two scissoring modes ($q_{4}$ and $q_{5}$) are degenerate. In partially
deuterated isotopologues all normal modes are nondegenerate.} \label%
{fig:nh3_vib_norm_modes}
\end{figure}

In the following, we only discuss the normal-mode evolution for the TGA,
which was the only one among the considered methods that reproduced the
experimental spectra accurately. Figure~\ref{fig:isotopologues_norm_mode_evo}
shows the evolution of all vibrational normal modes during the propagation.
As NH$_{3}$ and ND$_{3}$ possess the same symmetry (D$_{3h}$), the same
normal modes---symmetric stretching ($q_{3}$) and symmetric bending ($q_{6}$%
)---are activated. The symmetric stretching evolves with approximately twice
the frequency of the symmetric bending mode not only in the NH$_{3}$ and ND$%
_{3}$, but also in the partially deuterated isotopologues, suggesting that
the two modes are strongly coupled. Although the symmetric stretching is
always excited in all isotopologues, it appears to be absent in the spectra,
which has been further confirmed by jet-cooled experiments\cite%
{Vaida_Botschwina:1987, Syage_Steadman:1992} with NH$_{3}$ and ND$_{3}$. The
single progression in the spectra can be explained partially by the simple
fact that the bending mode is excited considerably more than other modes and
partially by invoking the missing mode effect (MIME).\cite%
{book_Heller:2018,Tutt_Zink:1982,Tutt_Heller:1987} In the MIME, the two
displaced modes collude at a time $t_{M}$ (with frequency $%
\omega_{M}=2\pi/t_{M}$), which here happens to correspond to the progression
of the symmetric bending mode. As there are more modes activated in
partially deuterated isotopologues, Table~\ref{tab:mime_freq} shows a
comparison between the observed experimental frequency and the calculated
MIME frequency, which indicates that in all cases the MIME is present and
its frequency corresponds to the symmetric bending mode. The additionally
excited vibrational normal modes in partially deuterated isotopologues are
one of the asymmetric stretching modes ($q_{1}$ or $q_{2}$ in Fig.~\ref%
{fig:isotopologues_norm_mode_evo}) and one scissoring normal mode ($q_{4}$
or $q_{5}$ in Fig.~\ref{fig:isotopologues_norm_mode_evo}). Interestingly,
one of the asymmetric stretching modes also evolves with approximately twice
the frequency of the bending mode, whereas the scissoring modes are
incommensurate with the rest of modes.

\begin{table}
\centering
\begin{tabular}{rllll}
& NH$_{3}$ & NDH$_{2}$ & ND$_{2}$H & ND$_{3}$ \\ \hline\hline
$\Delta \bar{\mu}$ & 935 & 868 & 781 & 704 \\ 
MIME frequency & 928 & 841 & 796 & 727%
\end{tabular}
\caption{Average peak spacing in the experimental spectra compared to the
MIME wavenumber calculated as $\Bar{\omega}= \sum_{j}\omega_{j}^{2}\delta
q_{j}^{2} / (\sum_{j}\omega_{j}\delta q_{j}^{2}N_{j})$, where
$N_{j}=[\omega_{j}/\Bar{\omega}]$ is the nearest integer to the indicated
ratio, $\omega_{j}$ is the wavenumber of the mode, and $\delta q_{j}$ is the
displacement in the mass-scaled normal-mode coordinates [see also Eq.~(19.2)
in Ref.~\citenum{book_Heller:2018}].} \label{tab:mime_freq}
\end{table}

\begin{figure}
\centering
\includegraphics{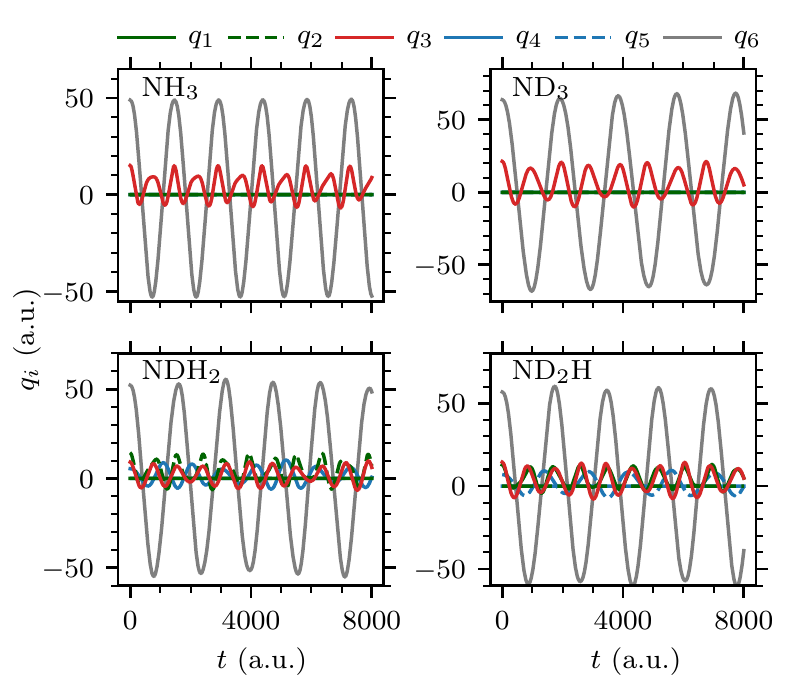} 
\caption{Evolution of the six
excited-state normal modes in the four isotopologues of ammonia following
electronic excitation at $t=0$. The normal mode coordinates $q_{1}$, $q_{2}$
correspond to the two asymmetric stretching modes, $q_{3}$ to the symmetric
stretch, $q_{4}$, $q_{5}$ to the two scissoring modes, and $q_{6}$ to the
symmetric bending (i.e., the umbrella vibration) about the minimum of the
excited-state PES (see also Fig.~\ref{fig:nh3_vib_norm_modes}).} \label%
{fig:isotopologues_norm_mode_evo}
\end{figure}

To conclude, we have shown that even the rather simple on-the-fly \textit{ab
initio} thawed Gaussian approximation can capture the key isotope effects in
the spectra of ammonia isotopologues. In contrast, the popular global
harmonic models can reproduce some of the isotope effects, but
inconsistently. The vertical harmonic model, where the PES is computed in
the Franck-Condon region, correctly describes the change in the width of the
spectral envelope but misses the isotope effect on the peak spacing. The
adiabatic harmonic model shows two progressions instead of the single
progression observed in experimental spectra. Inspection of the time
evolution of excited-state normal modes shows that the single progression in
the spectra can be explained by the larger excitation of the symmetric
bending mode than of the other modes and by the missing mode effect.\cite%
{book_Heller:2018, Tutt_Zink:1982} We also show that due to the change of
symmetry, in partially deuterated isotopologues additional modes are
activated, even though the symmetric bending mode still dominates the
dynamics and spectra.

\begin{acknowledgement}
The authors acknowledge the financial support from the European Research
Council (ERC) under the European Union's Horizon 2020 research and
innovation program (grant agreement No. 683069 -- MOLEQULE) and from the
EPFL.
\end{acknowledgement}

\begin{suppinfo}
Details of 1D harmonic model calculations, optimized geometries, harmonic frequencies, frequency shifts applied to computed spectra, normal mode evolution for global harmonic models, comparison between on-the-fly TGA and global harmonic model autocorrelation functions, and equation for the calculation of autocorrelation function in Hagedorn's parametrization.
\end{suppinfo}

\bibliography{Append_biblio53,biblio53}

\end{document}